\documentclass[12pt]{article}

\usepackage{times}
\usepackage{physics}
\usepackage{graphicx}
\usepackage{amssymb}
\usepackage{amsthm}
\usepackage{amsmath}
\usepackage{dsfont}
\usepackage{bm}
\usepackage{mathrsfs}
\usepackage{bbold}
\usepackage{color}
\usepackage{xcolor}
\usepackage{soul}
\usepackage[superscript]{cite}

\def\Xint#1{\mathchoice
   {\XXint\displaystyle\textstyle{#1}}%
   {\XXint\textstyle\scriptstyle{#1}}%
   {\XXint\scriptstyle\scriptscriptstyle{#1}}%
   {\XXint\scriptscriptstyle\scriptscriptstyle{#1}}%
   \!\int}
\def\XXint#1#2#3{{\setbox0=\hbox{$#1{#2#3}{\int}$}
     \vcenter{\hbox{$#2#3$}}\kern-.5\wd0}}

\def\dashint{\Xint-}

\begin{document}

\title{\bf Wigner function method for the Gibbons--Hawking and the Unruh effect}
\author{Ziv Landau and Ulf Leonhardt\\
Department of Physics of Complex Systems,\\
Weizmann Institute of Science,\\ Rehovot 7610001, Israel
}
\date{\today}
\maketitle

\begin{abstract}
An observer at rest with the expanding universe experiences some extra noise in the quantum vacuum, and so does an accelerated observer in a vacuum at rest  (in Minkowski space). The literature mainly focuses on the ideal cases of exponential expansion (de--Sitter space) or uniform acceleration (Rindler trajectories) or both, but the real cosmic expansion is non--exponential and real accelerations are non--uniform. Here we use the frequency--time Wigner function of vacuum correlations to define time--dependent spectra. We found excellent Planck spectra for a class of realistic cosmological models, but also strongly non--Planckian, negative Wigner functions for a standard scenario testable with laboratory analogues.\\

\end{abstract}

\newpage

\section{Introduction}

\begin{figure}[h]
\begin{center}
\includegraphics[width=20pc]{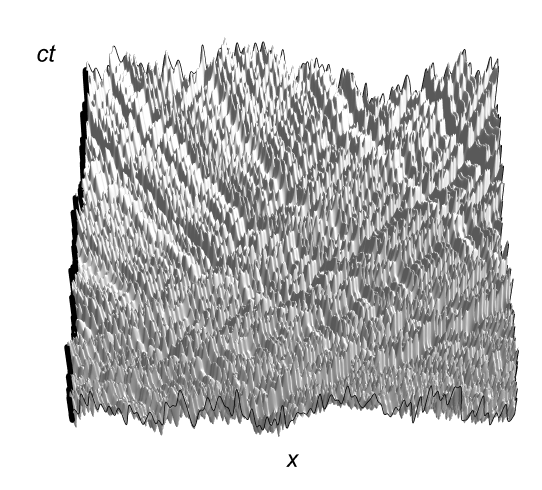}
\caption{
\small{Visualisation. Space--time diagram of vacuum noise in the expanding universe. An observer at rest samples the field (black curve on the left). Plot of 128 normalized modes summed up with Gaussian random complex amplitudes for the model (\ref{friedmann}) with $\gamma=3/2$ describing the current era of the universe (transition from matter to vacuum domination). One sees correlations in the noise --- curves of equal amplitudes --- along the light cones $s^2=0$  [Eq.~(\ref{s2}) with conformal time (\ref{tau}) given by Eq.~(\ref{tau1})].
}
\label{fig:noise}}
\end{center}
\end{figure}

Imagine an observer at rest with the empty space of the expanding universe (Fig.~\ref{fig:noise}). Suppose the observer is equipped with a sensitive spectrometer for detecting electromagnetic radiation. The spectrometer picks up the noise of vacuum fluctuations and also some extra noise that closely resembles the Hawking radiation of black holes \cite{Brout,GH}. The extra noise depends on the expansion, and it is due to a difference in the measure of time the observer and the vacuum perceive. The spectrometer decomposes the field into frequency components with respect to the proper time of the observer, whereas the fluctuations of the field (Fig.~\ref{fig:noise}) are continuously red--shifted by cosmic expansion \cite{Harrison}. A similar difference in time appears in the opposite scenario: an accelerated observer in a vacuum at rest \cite{Fulling,Davies,Unruh,Takagi} (in non--expanding Minkowski space). There the proper time of the observer differs from the time the vacuum experiences, too. The extra radiation is related to the Hawking radiation of black holes \cite{Brout}, because in both cases the observer is shrouded by horizons. In the expanding universe, it is the cosmological horizon \cite{Harrison} where the expansion velocity reaches the speed of light. In the case of the accelerated observer, it is the Rindler horizon \cite{Brout} where the observer approaches the speed of light. For an exponentially expanding, spatially flat universe (de--Sitter space) the cosmological horizon is light--like \cite{GH} as the event horizon of an ideal black hole \cite{Brout} and the observer would measure a perfect Planck spectrum \cite{GH}. But the real evolution of the universe is not exponential yet \cite{MC}. What is the spectrum there? The same question applies to the accelerated observer. A uniformly accelerated observer would measure a true Planck spectrum. But real accelerations must begin and end some time or may not be uniform at all. Is then the spectrum still Planckian? 

Renaud Parentani \cite{Macher1,Macher2,Finazzi} had worked out the precise condition when real black holes radiate a Planck spectrum. These real black holes are black--hole analogues in Bose--Einstein condensates \cite{Steinhauer}; they are real, because they are made in real experiments \cite{Steinhauer} and they correspond to a wide class of regimes \cite{Jacobson} likely to affect the emission of Hawking radiation from astrophysical black holes. In an ideal gravitational black hole, the speed of light does not depend on wavelength, even at the event horizon where light would stand still, oscillating with wavelengths shorter than any scale. For real astrophysical black holes, the wave velocity will eventually change with wavelength \cite{Jacobson}, releasing light from the grip of the horizon. In Bose--Einstein condensates, the analogue of the speed light --- the speed of sound --- definitely depends on the wavelength \cite{PS}. Analogues of cosmic expansion have been made as well and in those analogues particle creation was seen \cite{Viermann,Stein}. Almost arbitrary expansion histories of the universe can be tested \cite{FF1,FF2} with Bose--Einstein condensates  \cite{Viermann}. The related case of radiation perceived by non--uniform acceleration has been a notoriously difficult theoretical problem \cite{Milgrom,Obadia,KP,Barbado,Doukas,RousselFeller} but can be measured in analogues as well \cite{LeoUnruh,Gooding}.

In this paper, we hope to shed some light on the radiation spectrum in the expanding universe, known as the Gibbons--Hawking effect \cite{GH}, for non--exponential expansion and on the radiation perceived by accelerated observers, the Fulling--Davies--Unruh effect \cite{Fulling,Davies,Unruh,Takagi} or Unruh effect in short, for  non--uniform acceleration. The paper builds upon and corrects previous work \cite{RousselFeller,LeoEPL,LeoBerry} using the Wigner function \cite{Wigner,CahillGlauber,SchleichBook} for defining and calculating time--dependent spectra. We found excellent agreement with a Planck spectrum for one class of models, and poor agreement for another. What the condition for thermality and the effective temperatures are remains a mystery; it would probably take someone like Renaud Parentani to clarify it. 

\section{Correlation function}

Consider a spatially flat universe expanding with scale factor $a(t)$ where $t$ denotes the time perceived by an observer at rest, the cosmological time. The quantum vacuum of the electromagnetic field propagates with conformal time \cite{MC} (Appendix A):
\begin{equation}
\tau = \int \frac{\mathrm{d}t}{a} \,.
\label{tau}
\end{equation}
The difference between the time $t$ of the observer and the time $\tau$ of the vacuum is the key to the Gibbons--Hawking effect \cite{LeoBerry}.Consider the correlation function of the electromagnetic vector potential at the two space--time points $\{t_1,{\bm r}_1\}$ and $\{t_2,{\bm r}_2\}$  in co--moving coordinates \cite{MC}.In an isotropic and homogeneous universe \cite{MC}, the two electromagnetic polarizations are independent from each other \cite{LeoLambda} and have equal correlation functions:
\begin{equation}
K_0 = \frac{\varepsilon_0 c}{2\hbar} \langle 0| \widehat{A}_1 \widehat{A}_2 + \widehat{A}_2 \widehat{A}_1|0\rangle 
\label{Kdef}
\end{equation}
where $\varepsilon_0$ denotes the electric permittivity of the vacuum ($c$ the speed of light and $\hbar$ the reduced Planck constant) while the indices in $ \widehat{A}_1$ and $ \widehat{A}_2$ refer to the space--time points $\{t_1,{\bm r}_1\}$ and $\{t_2,{\bm r}_2\}$ the field operators are evaluated at. The vacuum energy density and pressure of the field can be calculated from $K_0$ by differentiating with respect to the space--time coordinates. The correlation $K_0$ is the real part of the analytic function $(\varepsilon_0 c/\hbar)\langle 0| \widehat{A}_1 \widehat{A}_2 |0\rangle$ while its imaginary part corresponds to the dissipation \cite{LeoLambda}:
\begin{equation}
\Gamma_0 = \frac{\varepsilon_0 c}{2\mathrm{i}\hbar} \, \big[\widehat{A}_2,\widehat{A}_1\big] = \frac{1}{2c} (G_+-G_-) 
\label{K}
\end{equation}
where the $G_\pm$ are the retarded and advanced Green functions. These Green functions are entirely classical and independent of the quantum state. The advanced Green function $G_-$ describes a wave focusing at a point drain with unit strength, while the retarded Green function $G_+$ describes the wave emitted from a point source of unit strength. The difference between the advanced and the retarded Green function accounts for the dissipation. For analytic functions $f$ with 
\begin{equation}
K = \mathrm{Re} f \,,\quad \Gamma = \mathrm{Im} f 
\label{FDT}
\end{equation}
the correlation $K$ of the fluctuations is related to the dissipation $\Gamma$. This is the  fluctuation--dissipation theorem \cite{LeoLambda}. Note that the correlation $K_0$ depends on the quantum state, the vacuum state in our case, while the dissipation is state--independent. For the electromagnetic vacuum, we encode the state requiring that the Fourier transform of $f_0$ with respect to conformal time (\ref{tau}) contains only positive frequencies. This is the condition of the {\it conformal vacuum} and it comes from the fact that the modes of the electromagnetic field are waves of positive frequencies with respect to $\tau$. From this condition  follows the analyticity and decay of $(\varepsilon_0 c/\hbar)\langle 0| \widehat{A}_1 \widehat{A}_2 |0\rangle$ on the lower half $\tau$ plane and hence the Kramers--Kronig relation (known in mathematics as the Hilbert transformation):
\begin{equation}
K_0(\tau) = -\frac{1}{\pi} \dashint_{-\infty}^{+\infty} \frac{\Gamma_0(\tau_0)}{\tau_0-\tau}\,\mathrm{d}\tau_0  \,.
\label{KK}
\end{equation}
In a spatially flat universe, electromagnetic fields propagate with respect to conformal time (\ref{tau}) like in Minkowski space \cite{MC}. We thus have $G_\pm=-(4\pi r)^{-1}\delta(\tau_{21}\mp r/c)$ with $\tau_{21}=\tau_2-\tau_1$ and $r=|{\bm r}_2-{\bm r}_1|$.  The Kramers--Kronig relation (\ref{KK}) turns each delta function in the dissipation to a pole in the correlation:
\begin{equation}
K_0 =  - \frac{1}{8\pi^2 r}\left(\frac{1}{c\tau_{21}-r}-\frac{1}{c\tau_{21}+r}\right) .
\label{K0}
\end{equation}
This is the correlation function of the electromagnetic vacuum fluctuations (Fig.~\ref{fig:noise}). Note that such a correlation function has been measured inside a dielectric material \cite{ETH}. If we divide the vector--potential component $A$ of each polarization by the scale factor $a$ we obtain the conformally--coupled scalar field $\phi$ (Appendix A). It is the correlations of this field we are considering here: 
\begin{equation}
K =  \frac{K_0}{a_1 a_2} = - \frac{1}{(2\pi)^2 s^2}
\label{K}
\end{equation}
where we get from Eq.~(\ref{K0}):
\begin{equation}
s^2 = a_1 a_2 \left[c^2(\tau_2-\tau_1)^2-({\bm r}_2-{\bm r}_1)^2\right] .
\label{s2}
\end{equation}
We treat the observer as a point detector at rest, ${\bm r}_1={\bm r}_2=\mathrm{const}$, sampling the correlation function $K$ along the proper time of the observer, the cosmological time $t$. We thus obtain from Eqs.~(\ref{tau}), (\ref{K}) and (\ref{s2}):
\begin{equation}
K = - \frac{1}{(2\pi)^2c^2a_1a_2(\tau_2-\tau_1)^2}  = - \frac{1}{(2\pi)^2c^2} \, \frac{\dot{\tau}_2\,\dot{\tau}_1}{(\tau_2-\tau_1)^2} 
\label{Ktau}
\end{equation}
with the dots denoting differentiation  with respect to cosmological time. If we replace $\tau$ by the M\"{o}bius transformation 
\begin{equation}
\tau\rightarrow \frac{\alpha\,\tau+\beta}{\gamma\,\tau+\delta}
\label{moebius}
\end{equation}
with constant  coefficients $\alpha$, $\beta$, $\gamma$, $\delta$ the correlation function (\ref{Ktau}) remains unchanged.Note that this feature is not related to M\"{o}bius transformations of the cosmological time $t$ studied elsewhere \cite{BALivine1,BALivine2} (despite using there the notation $\tau$ for $t$).

As an example, take the standard case of exponentially expanding space, de--Sitter space. There we have $a=a_0\mathrm{e}^{H t}$ with Hubble constant $H$ and get from Eq.~(\ref{tau}) the conformal time $\tau=-(aH)^{-1}$. This gives for $K$ formula (\ref{K}) with
\begin{equation}
s^2 = \frac{c^2}{H^2} \,4\sinh^2\frac{H}{2}\theta\quad\mbox{and}\quad \theta=t_2-t_1 \,.
\label{deS}
\end{equation}
Consider now the thermal state in Minkowski space ($a=1$) with temperature $T$ written as 
\begin{equation}
k_\mathrm{B} T = \frac{\hbar H}{2\pi} \,.
\label{gh}
\end{equation}
One gets from the fluctuation--dissipation theorem (\ref{FDT}) with Kubo--Martin--Schwinger relation \cite{LeoLambda} the correlation function:
\begin{equation}
K_\mathrm{th} = - \frac{1}{8\pi^2 c^2\rho^2}\,\partial_\rho \ln\left(\mathrm{e}^{H\theta}-\mathrm{a}^{H\rho}\right)\left(\mathrm{e}^{H\theta}-\mathrm{a}^{-H\rho}\right)
\quad\mbox{with}\quad \rho=\frac{|{\bm r}_2-{\bm r}_1|}{c} \,.
\end{equation}
One verifies that $K$ in de--Sitter space [Eqs.~(\ref{K}) and (\ref{deS})] agrees with $K_\mathrm{th}$ in Minkowski space for ${\bm r}_1={\bm r}_2$. This shows that the observer at rest in de--Sitter space perceives the quantum vacuum as a thermal state with Gibbons--Hawking temperature (\ref{gh}). Note that it has been essential here to treat the observer as a point detector, as the correlation function $K$ differs from $K_\mathrm{th}$ for ${\bm r}_1\neq{\bm r}_2$. Nonlocal detectors will respond differently to cosmic expansion. For example, one may take the ionisation of an atom by the expansion of space as a detection mechanism for Gibbons--Hawking radiation \cite{Volovik}. As the released electron tunnels out of the atom to infinity, this is not a local detection. For exponential expansion one gets exactly twice \cite{Volovik} the Gibbons--Hawking temperature of Eq.~(\ref{gh}) --- presumably because the expansion does not only facilitate but also accelerate the escape of the electron.

Finally, we turn from the Gibbons--Hawking effect of the observer at rest with the expanding universe to the Unruh effect of the accelerated observer. In the Unruh effect \cite{Fulling,Davies,Unruh,Takagi} a detector moves in empty Minkowski space on the space--time trajectory $t(\tau)$ and ${\bm r}(\tau)$ parameterized by the proper time of the detector, $\tau=\int \mathrm{d}\tau$ with $\mathrm{d}\tau^2=\mathrm{d}t^2-\mathrm{d}{\bm r}^2/c^2$. Here the roles of $t$ and $\tau$ are reversed: the vacuum propagates with Minkowski time $t$ while the spectrometer of the detector responds to the proper time $\tau$. The detector thus samples the correlation function (\ref{K}) with 
\begin{equation}
s^2 = \left[c^2(t_2-t_1)^2-({\bm r}_2-{\bm r}_1)^2\right] .
\label{sunruh}
\end{equation}
A uniformly moving detector would just measure vacuum fluctuations, as $s^2$ is invariant under Lorentz transformations, but an accelerated observer would see some additional noise. In the case of uniform acceleration we have \cite{Brout} $ct=\xi\sinh\eta$, $x=\xi\cosh\eta$, $y=z=0$ for $\xi=\mathrm{const}$. The acceleration is given by $a=c^2/\xi$ while the proper time is described by the dimensionless parameter $\eta$ as $\tau=(\xi/c)\eta$. We obtain $s^2=\xi^2\,4\sinh^2(\eta_2/2-\eta_1/2)$, which corresponds to the de--Sitter correlation (\ref{deS}) and hence describes thermal radiation as well. Note that this result depends on the number of spatial dimensions (in our case 3) and the spin of the field (for the electromagnetic field spin 1) \cite{Takagi}.

\section{Wigner function}

The spectrometer of the observer does two things: it decomposes the amplitude into spectral components and it detects them. As the spectrometer is attached to the observer, the spectrum is taken with respect to the proper time of the observer. This, in our opinion, is the most important aspect of the Gibbons--Hawking and the Unruh effect, not the actual act of detection \cite{LeoUnruh}. It does not matter what exactly the detector does --- it could measure the amplitude or count radiation quanta --- as long as it is a point--detector. The spectrum is assumed to be continuously measured during the cosmic expansion or during the accelerated motion of the observer. For exponential expansion or uniform acceleration the correlation function depends only on the time difference. We could therefore describe the power spectrum by the Fourier transform with respect to that time difference. But in general, the correlation function will depend on two independent times. We could still define a time--dependent spectrum by Fourier--transforming with respect to the time difference:
\begin{equation}
W = \frac{1}{2\pi} \int_{-\infty}^{+\infty} \mathrm{e}^{\mathrm{i}\omega\theta} K(t+\theta/2,t-\theta/2)\,\mathrm{d}\theta \,.
\label{wigner}
\end{equation}
This is the chronocyclic Wigner function adapted from classical \cite{Paye} to quantum fields, the central subject of this paper. It was introduced before \cite{RousselFeller, LeoBerry}, but in one study \cite{RousselFeller} it would produce highly oscillatory spectra when applied to cosmology and in the other \cite{LeoBerry} it was applied incorrectly for working out the high--frequency asymptotics of the spectrum. Here we will argue that it is a useful tool for low frequencies. 

The Wigner function \cite{Wigner,CahillGlauber,SchleichBook} was originally introduced \cite{Wigner} as a quasiprobability distribution for the position and the momentum of a quantum particle (and first applied in thermodynamics \cite{Wigner}). It can easily be generalized to other Fourier conjugates such as frequency and time, it has been generalized to angular--momentum variables \cite{Angular} and to discrete variables \cite{LeoCheese1,LeoCheese2}, and it is routinely measured in quantum state tomography \cite{Smithey,Raymer,LeoMeasuring}. For canonically conjugate variables, such as frequency and time, no proper phase--space distribution can exist, but the Wigner function acts as a good compromise. Its marginal distributions do give the correct power spectra or intensity distributions under linear transformations of the conjugate variables. This is its defining property \cite{LeoMeasuring}. 

\begin{figure}[h]
\begin{center}
\includegraphics[width=20pc]{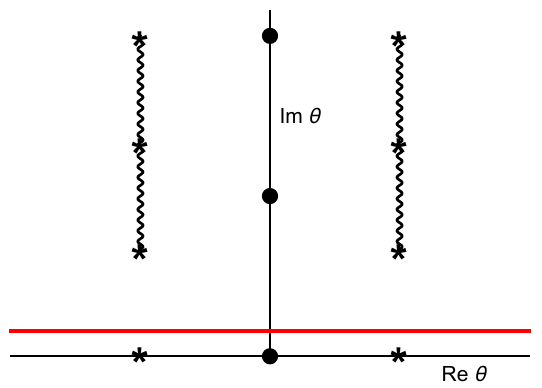}
\caption{
\small{Contour. Amended integration contour of the Wigner function (\ref{wigner}) for complex time difference $\theta=t_2-t_1$ . For $\omega>0$ the contour (red) should lie above the real axis avoiding the singularities (dots) and branch cuts (zigzags) between the branch points (stars) of the correlation function $K$. For $\omega<0$ the contour lies below the real axis. The figure corresponds to the model of Eq.~(\ref{friedmann}) with $\gamma=3/2$ describing the current cosmological era (as in Fig.~\ref{fig:noise}). For other $\gamma>1$ the structure of $K$ is similar. 
}
\label{fig:contour}}
\end{center}
\end{figure}

In our case, the correlation function $K$ given by Eq.~(\ref{Ktau}) is an even, real function with respect to the time difference $\theta=t_2-t_1$. From this follows that the Wigner function (\ref{wigner}) is real and identical for $\pm\omega$. The same is true for the Wigner function of the accelerated observer. For an exponentially expanding universe (de--Sitter space) we get \cite{RousselFeller}: 
\begin{equation}
W = \frac{1}{(2\pi)^2c^2} \left(\frac{\omega}{2}+\frac{\omega}{\mathrm{e}^{2\pi\omega/H}-1}\right) \quad \mbox{for}\quad \omega>0
\label{deSwigner}
\end{equation}
independent of time $t$, and the same expression with $\omega$ replaced by $|\omega|$ for $\omega<0$. We may interpret this Wigner function as being proportional to the thermal energy of a bosonic point particle: the $\omega/2$ term describes the zero--point energy and the other term the Planck distribution of a thermal state with temperature (\ref{gh}). The point detector acts as a point particle in thermal equilibrium with Gibbons--Hawking temperature (\ref{gh}). Mathematically, the zero--point term appears due to the double pole of the correlation function (\ref{Ktau}) at $\theta=0$:applying for $r\neq 0$ Cauchy's theorem through the single poles on the real $\theta$ axis and taking the limit $r\rightarrow 0$ produces this term. This is true not only for exponential expansion, but for any $a(t)$, which is consistent with the physical picture of this term describing the vacuum noise floor of the detector. We may remove the vacuum term by moving the integration contour in formula (\ref{wigner}) infinitesimally above the real axis (Fig.~\ref{fig:contour}) for $\omega>0$ (or below it for $\omega<0$) as this contour does not cross the poles and hence does not produce the vacuum term.

Moving the integration contour has another advantage \cite{LeoBerry}. At the beginning of time, say at $t=0$, the scale factor must develop a branch point \cite{LL2} for a realistic universe such that $a$ is no longer real or positive for $t<0$ --- there was no discernible universe before the beginning of time. This would restrict the integration interval in the Wigner function (\ref{wigner}) to $[-2t,+2t]$. However, we may analytically continue the correlation function $K$ beyond the physically relevant interval $[-2t,+2t]$. As the spectrum should primarily depend on cosmologically short times around $t$ this extension should not affect it much. But there is a problem. Suppose that $a\propto t^{1/\gamma}$ with constant $\gamma\ge 1$ for $t\sim 0$. From this follows [Eq.~(\ref{Ktau})] that $K\propto (t+\theta/2)^{-1/\gamma}+(t-\theta/2)^{-1/\gamma}$ as leading contribution near $\theta=\pm 2t$. We obtain from Wigner's formula (\ref{wigner}):
\begin{equation}
W \propto \frac{\cos\left(2t\omega-\frac{\pi}{2\gamma}\right)}{\omega^{1-1/\gamma}}\quad\mbox{for}\quad\gamma\ge 1 \,.
\label{wignerosc}
\end{equation}
We thus get spectral oscillations. Fourier transformations are known to generate oscillations near sharp features; in our casethese originate from the branch points in the integrand of the Wigner function. Moving the integration contour slightly above or below the real axis avoids the branch points (Fig.~\ref{fig:contour}) and thus removes this artefact. In a previous study \cite{RousselFeller} the integration was left on the real axis, which would have produced highly oscillatory, meaningless spectra if they were actually calculated, whereas for our integration contour we may get physically meaningful spectra of the extra noise due to cosmic expansion. 

Suppose, hypothetically, the universe were filled with one ideal fluid of energy density $\varepsilon$ and pressure $p$ with the equation of state $p=w\varepsilon$ and $w=\mathrm{const}$. From thermodynamics follows \cite{LL2} $\mathrm{d}\varepsilon/\mathrm{d}a=-3(\varepsilon+p)/a=-3(1+w)\varepsilon/a$. Solving this differential equation gives $\varepsilon\propto a^{-2\gamma}$ with 
\begin{equation}
\gamma = \frac{3}{2}\,(1+w) \,.
\label{gamma}
\end{equation}
From the Friedmann equation \cite{LL2} follows that $a\propto t^{1/\gamma}$. This is the case considered above. The amended Wigner function would be exactly zero, which is consistent with summing up all the phases in a cascade of Gibbons--Hawking processes at cosmological horizons \cite{LeoBerry}. They also give net zero radiation for a universe filled with a single fluid. 

Consider now the realistic case of a fluid plus the cosmological constant (often called ``the vacuum''). In this case the Friedmann equation \cite{LL2} reads
\begin{equation}
H^2 = H_\infty^2\left(1+a^{-2\gamma}\right)
\label{friedmann}
\end{equation}
with $\gamma$ given by Eq.~(\ref{gamma}) and $H$ being the Hubble parameter
\begin{equation}
H = \frac{\dot{a}}{a} \,.
\label{hubble}
\end{equation}
The constant $H_\infty$ amounts to the Hubble parameter infinitely in the future when $a\rightarrow\infty$. For $w=0$ and hence $\gamma=3/2$ this model describes the current era of the universe turning from matter to vacuum domination. We obtain from Eq.~(\ref{tau}) for the conformal time and definition (\ref{hubble}) of the Hubble parameter:
\begin{equation}
t = \int \frac{\mathrm{d}a}{aH} \quad \implies \quad a = \left(\sinh\gamma H_\infty t\right)^{1/\gamma} \,.
\label{t}
\end{equation}
In the special case of $\gamma=1/2$ the scale factor turns into the square of the $\sinh$ of $H_\infty t/2$ and so the conformal time [Eq.~(\ref{tau})] becomes $\tau=-2H_\infty ^{-1}\coth (H_\infty t/2)$. This is a M\"{o}bius transformation (\ref{moebius}) of the conformal time for exponential expansion, $\tau=-H_\infty^{-1}\mathrm{e}^{-H_\infty t}$. We thus conclude that the Wigner function gives a perfect Planck spectrum for $\gamma=1/2$. For general $\gamma$ we obtain from the definition (\ref{tau}) of conformal time
\begin{equation}
\tau = \int \frac{\mathrm{d}a}{a^2H} = H_\infty^{-1} \,\frac{a^{\gamma-1}}{\gamma-1}\, {}_2F_1 \left(\frac{1}{2},\frac{1}{2}-\frac{1}{2\gamma},\frac{3}{2}-\frac{1}{2\gamma},-a^{2\gamma}\right)
\label{tau1}
\end{equation}
in terms of Gauss' hypergeometric function ${}_2 F_1$. We calculated numerically the Wigner function with the amended integration contour (Fig.~\ref{fig:contour}) for a range of $\gamma$ parameters (with $\gamma>1$) and times $t$, and found excellent agreement with Planck spectra for low frequencies.

\begin{figure}[h]
\begin{center}
\includegraphics[width=20pc]{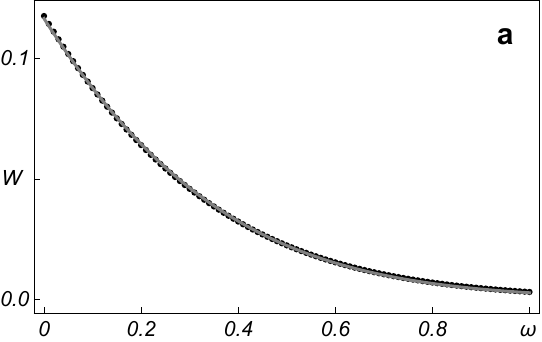}
\includegraphics[width=20pc]{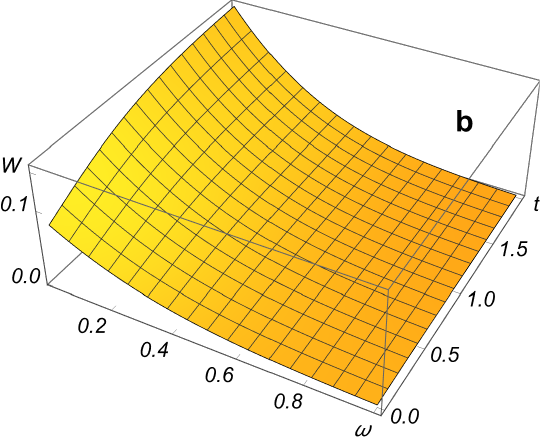}
\caption{
\small{Planck spectra. Wigner function (\ref{wigner}) [in units of $1/(2\pi)^2c^2$] with integration contour (Fig.~\ref{fig:contour}) calculated numerically for the model (\ref{friedmann}) with $\gamma=3/2$. Frequency $\omega$ and time $t$ are given in units of $H_\infty$ and $1/H_\infty$, respectively. For different $\gamma>1$ we get similar results. {\bf a} shows the numerical Wigner function (dots) for $H_\infty t = 0.5$, a characteristic time during the transition from matter to vacuum domination. The Planckian fit (gray curve) of Eq.~(\ref{fit}) agrees well with the data for $H_\mathrm{eff}=1.155 H_\infty$ and prefactor $A=0.6327$. {\bf b}: Wigner function for a range of frequencies and times. One sees that the Planck spectrum hardly varies in time. 
}
\label{fig:planck}}
\end{center}
\end{figure}

Figure \ref{fig:planck} shows the results for the physically most relevant case, $\gamma=3/2$ (the current era of transition from matter to vacuum domination). The numerical data points agree with a Planckian fit well into the exponential tail of the spectrum (Fig.~\ref{fig:planck}a). We use as fit
\begin{equation}
W=\frac{A}{(2\pi)^2c^2} \,\frac{\omega}{\mathrm{e}^{2\pi\omega/H_\mathrm{eff}}-1} \,.
\label{fit}
\end{equation}
characterized by the effective Hubble parameter $H_\mathrm{eff}$ for $H$ in the Gibbons--Hawking temperature (\ref{gh}) and the prefactor $A$. The effective temperature depends on time, but it does not vary much. If the spectra would vary rapidly in time, one could question whether they do represent actual spectra. Let us estimate this. The characteristic time $\Delta t$ over which the spectrum (\ref{fit}) varies is given by $1/\Delta t\sim|\partial_t H_\mathrm{eff}|/H_\mathrm{eff}$. The characteristic variation $\Delta\omega$ of the frequency in the Planck spectrum (\ref{fit}) is $\Delta\omega\sim H_\mathrm{eff}$. The spectrum would be affected by frequency--time uncertainty if $\Delta t\,\Delta\omega\sim 1$. However, our numerical results show that $|\partial H_\mathrm{eff}^{-1}/\partial t| \ll 1$ by about $10^{-2}$ and so $\Delta t\,\Delta\omega\gg 1$. The Planck spectra captured by the Wigner function (Fig.~\ref{fig:planck}) are genuine for low frequencies. 

\begin{figure}[h]
\begin{center}
\includegraphics[width=20pc]{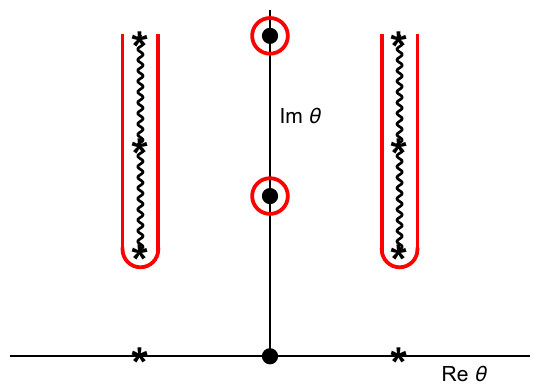}
\caption{
\small{Contour deformation. The integration contour (red) of the Wigner function (Fig.~\ref{fig:contour}) is deformed to go around the singularities (dots) and branch cuts (zigzags). For high frequencies the Fourier factor $\mathrm{e}^{\mathrm{i}\omega\theta}$ in the integral (\ref{wigner}) decays rapidly on the upper half plane. Above the real axis, the lowest branch point lies lower than the lowest singularity and therefore dominates the spectrum. 
}
\label{fig:deform}}
\end{center}
\end{figure}

For high frequencies, the spectrum becomes exponentially decaying and oscillatory. This is a remnant of the oscillations due to the branch point at the beginning of physical time, Eq.~(\ref{wignerosc}). By moving the integration contour away from the real axis we have eliminated this feature in the main part of the spectrum, but it reappears in the exponential tail. Figure \ref{fig:deform} shows why. The integration contour can be deformed to go around the singularities and the branch cuts. The latter are situated on the lines $\mathrm{Re}\,\theta=\pm 2t$, $\mathrm{Im}\,\theta\ge 2\pi/\gamma$. The singularities produce, via Cauchy's theorem, a geometrical series that sums up to a Planck spectrum with $H=H_\infty$:
\begin{equation}
W_\infty = \frac{\omega}{(2\pi)^2c^2} \sum_{m=1}^\infty\mathrm{e}^{-2\pi \omega/H_\infty} = \frac{1}{(2\pi)^2c^2} \,\frac{\omega}{\mathrm{e}^{2\pi\omega/H_\infty}-1} \,.
\label{planckseries}
\end{equation}
The branch cuts generate the corrections $W_\mathrm{cuts}$ to $W_\infty$ (as $W=W_\infty+W_\mathrm{cuts}$). For high frequencies and complex $\theta$ the Fourier factor $\mathrm{e}^{\mathrm{i}\omega\theta}$ in the Wigner integral (\ref{wigner}) becomes exponentially decaying. For $\gamma>1$ the lowest branch point lies lower than the lowest singularity and so dominates the spectrum. Its imaginary part, $2\pi/\gamma$, gives an exponential decay, and its real part, $2t$, an oscillation. These features appear to be artefacts indicating the limitations of our Wigner method, but they only affect the exponential tail of the spectrum (Fig.~\ref{fig:planck}).

\begin{figure}[h]
\begin{center}
\includegraphics[width=18pc]{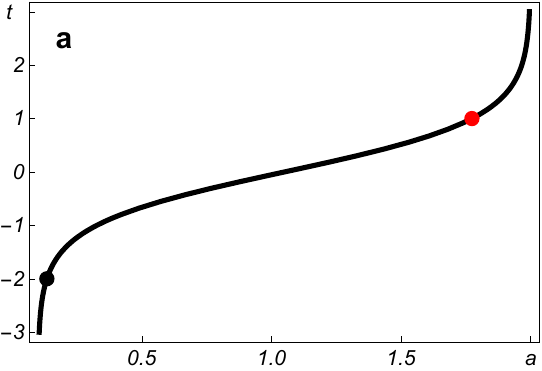}
\includegraphics[width=18pc]{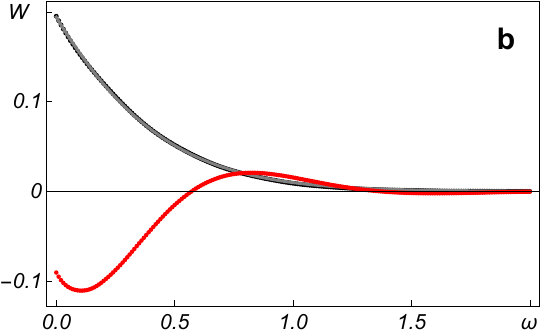}
\caption{
\small{Non--Planckian example. {\bf a}: Space--time diagram of the analogue of the scale factor $a$ evolving in time according to Eq.~(\ref{example}). The Wigner function (\ref{wigner}) is numerically calculated for two times (black and red dot) and shown in {\bf b}. For $H_0 t = -2$ (black) one gets a good Planck spectrum, but for $H_0 t = 1$ the Wigner function is non--Planckian and partly negative. 
}
\label{fig:example}}
\end{center}
\end{figure}

Let us give another example showing a peculiar feature of the Wigner function (Fig.~\ref{fig:example}). Consider an analogue of cosmic expansion where the scale factor $a$ turns from a near--constant $a_0$ to another near--constant $a_1$ (Fig.~\ref{fig:example}a). For example, suppose
\begin{equation}
a = \frac{a_0\, \mathrm{e}^{-H_0t} + a_1\, \mathrm{e}^{H_0 t}}{\mathrm{e}^{-H_0 t}+\mathrm{e}^{H_0 t}}
\label{example}
\end{equation}
with $H_0=\mathrm{const}$. In this case, we obtain for the conformal time (\ref{tau}):
\begin{equation}
\tau = \frac{(a_0+a_1)H_0t+(a_0-a_1)\ln(a_0\, \mathrm{e}^{-H_0t} + a_1\, \mathrm{e}^{H_0t})}{2a_0a_1 H_0} \,.
\label{tauexample}
\end{equation} 

\noindent For $H_0t\ll -1$ the conformal time approaches $t/a_0$ and for $H_0t\gg +1$ it tends to $t/a_1$, both as it should for constant scale factors $a$, according to definition (\ref{tau}). Figure \ref{fig:example}b shows the Wigner function. For $H_0t\lesssim -1$ we get a good Planck spectrum, but for $H_0t\gtrsim +1$ the Wigner function becomes negative. In quantum optics \cite{SchleichBook}, negative regions of the Wigner function indicate nonclassical features \cite{LeoMeasuring}; we are not sure what they mean here. 

\section{Conclusions}

We applied an amended \cite{LeoBerry} Wigner function \cite{Wigner,CahillGlauber,SchleichBook} for calculating the spectrum of vacuum fluctuations for an observer at rest in the expanding universe or, equivalently, for an accelerated observer in a space at rest (in empty Minkowski space). The literature mostly focuses on exponential expansion (de--Sitter space) and uniform acceleration (Rindler trajectories) because these are the ideal cases where the cosmological or the Rindler horizons are genuine event horizons. Here we have developed tools for addressing the real cases of non--exponential expansion and non--uniform acceleration. The key feature that allows the calculation of physically meaningful spectra is the modification of the integration contour in the Wigner function \cite{LeoBerry}. Otherwise the spectra would be highly oscillatory and meaningless. For a class of models that includes the model describing the current cosmological era, we found excellent fits with Planck spectra. The vacuum noise in the expanding universe appears as thermal noise. However, we were not able to derive a simple analytic formula for the radiation temperature; a previous formula \cite{LeoBerry} for the high--frequency limit turned out to be incorrect. For high frequencies in the exponential tail of the spectrum, the Wigner function exhibits spurious oscillations, and the Wigner method seems to fail there. For another type of cosmic evolution that can be experimentally tested in laboratory analogues \cite{Viermann,Stein,FF1,FF2} we found non--Planckian, negative Wigner functions. In quantum optics \cite{SchleichBook} negative Wigner functions indicate non--classical features. What are they in our case? When is the spectrum Planckian? What is the temperature? When science answers one question, many more questions arise. We leave it to future readers to find some of the answers.

\section*{Acknowledgements}

U.L. met Renaud Parentani on Capri in 2000, at a meeting on Cosmology in the Laboratory. While sitting together in the shade or strolling along the many paths on this beautiful island, talking physics, he learned from Renaud the beginning of the story that still occupies him today, the quantum physics of the cosmos. Many more meetings and discussions with Renaud were to follow, until a final one in Paris. His clarity and wit are missed now. We thank him for the gift he gave, and we also thank
Michael Berry,
Mathias Fink,
Uwe Fischer,
Amaury Micheli,
Scott Robertson,
Grisha Volovik
and
Chris Westbrook
for discussions that contributed to this paper. The paper was supported by the Murray B. Koffler Professorial Chair at the Weizmann Institute.

\appendix

\section{Appendix}
\renewcommand{\theequation}{\thesection.\arabic{equation}}
\setcounter{equation}{0}

In this appendix we present an elementary derivation of the relationship between electromagnetic waves and conformally coupled scalar fields \cite{BD} propagating in a spatially flat, expanding universe. The line element of the space--time metric, the Robertson--Walker--Friedmann--Lema\^{i}tre metric, reads \cite{LL2}:
\begin{equation}
\mathrm{d}s^2 = c^2\mathrm{d}t^2-a^2\mathrm{d}\bm{r}^2 \,.
\label{RWFL}
\end{equation}
For light rays $\mathrm{d}s=0$, and so we may divide the metric (\ref{RWFL}) by $a^2$ with the result
\begin{equation}
\mathrm{d}s^2_\mathrm{em} = \frac{c^2}{a^2}\, \mathrm{d}t^2-\mathrm{d}\bm{r}^2 =  c^2\mathrm{d}\tau^2-a^2\mathrm{d}\bm{r}^2
\label{conformal}
\end{equation}
according to the definition (\ref{tau}) of the conformal time $\tau$. As Maxwell's equations are conformally invariant \cite{BD} the metric (\ref{conformal}) remains valid for full electromagnetic waves and their fluctuations. Each component $A$ of the vector potential  thus satisfies the same wave equation in conformal time $\tau$ and comoving coordinates $\bm{r}$  as in Minkowski space. There we have in Coulomb gauge:
\begin{equation}
0 = \left(\partial_\tau^2 - c^2\nabla^2\right)A = \left(a\partial_t a\partial_t - c^2\nabla^2\right)A \,.
\label{waveeq}
\end{equation}
Let us define the differential operator
\begin{equation}
\Box = \frac{1}{c^2}\left(\partial_t^2+3H\partial_t\right) - \frac{\nabla^2}{a^2}
\label{dalembertian}
\end{equation}
in terms of the Hubble parameter (\ref{hubble}), and also the quantity
\begin{equation}
R = -\frac{6}{c^2}\,(\dot{H}+2H^2)
\label{riemann}
\end{equation}
and the field 
\begin{equation}
\phi=\frac{A}{a} \,.
\label{phi}
\end{equation}
From definitions (\ref{dalembertian}-\ref{phi}) follows by elementary calculations
\begin{equation}
\bigg(\Box - \frac{R}{6} \bigg) \phi = \frac{1}{a^3 c^2} \bigg(a\partial_t a\partial_t - c^2\nabla^2\bigg)A = 0
\label{conf}
\end{equation}
according to the wave equation (\ref{waveeq}). The important point of this relationship is the geometrical meaning of $\Box$ and $R$: $\Box$ is the d'Alembertian \cite{LL2}  $D_\alpha D^\alpha=\frac{1}{\sqrt{-g}}\partial_\alpha \sqrt{-g}g^{\alpha\beta}\partial_\beta$ with respect to the Robertson--Walker--Friedmann--Lema\^{i}tre metric (\ref{RWFL}) with $g=-a^6$ and $g^{\alpha\beta}=\mathrm{diag}(1,-a^{-2},-a^{-2},-a^{-2})$. One verifies that $R$ is the corresponding curvature scalar \cite{LL2}. The resulting wave equation (\ref{conf}) is the equation of a conformally coupled scalar field \cite{BD}. Q.~E.~D.

\newpage

\newpage

\newpage

\newpage

\end{document}